\documentclass[english]{article}
\usepackage[T1]{fontenc}
\usepackage[latin1]{inputenc}
\usepackage{graphicx}
\usepackage{amssymb}

\usepackage{color}

\makeatother
\begin{document}

\title{Uncertainty relations for the realisation of macroscopic quantum
superpositions and EPR paradoxes\\
 }

\author{E. G. Cavalcanti and M. D. Reid\\
 ARC Centre for Excellence for Atom-Optics,\\ School of Physical Sciences,
The University of Queensland, \\Brisbane, Australia\\
 }

\date{\today{}}

\maketitle
\begin{abstract}
We present a unified approach, based on the use of quantum uncertainty
relations, for arriving at criteria for the demonstration of the EPR
paradox and macroscopic superpositions. We suggest to view each criterion
as a means to demonstrate an EPR-type paradox, where there is an inconsistency
between the assumptions of a form of realism, either macroscopic realism
(MR) or local realism (LR), and the completeness of quantum mechanics. 
\end{abstract}

\section{Introduction}

Schrödinger \cite{sch} raised the question of whether there could
be a superposition of macroscopically distinct states. The issue at
hand\cite{leggett} is that where we have a quantum superposition
of two states, the system cannot be thought of as being in one state
or the other until a measurement is performed that would distinguish
the states.

The concept of the quantum superposition is intrinsically associated
with the concept of a \emph{fundamental quantum indeterminateness},
that we are limited in the precision to which we can ever predict
outcomes of measurements that are performed on the system. This follows
because if we have a superposition of two eigenstates $|x_1\rangle$ and $|x_2\rangle$ of an observable
$\hat{x}$, where $x_{2}-x_{1}$ is large, then by our interpretation, the
system is not predetermined to be in either state, so we have an indeterminacy
in the outcome $x$ that is at least of order $x_{2}-x_{1}$.

This indeterminacy is of a fundamentally different nature to that
of classical theory, where lack of knowledge of an outcome is understood
in terms of a statistical theory in which there is a probability for
the system to be in a certain state, which will have a certain probability
of outcome for $x$. Such probabilistic interpretations are generally
referred to as \emph{classical mixtures}. In quantum mechanics, the
indeterminacy that arises from a quantum superposition is not represented
this way.

The concept of a \emph{macroscopic superposition} is therefore linked
with that of a \emph{macroscopic} \emph{quantum} \emph{indeterminateness},
which manifests as a macroscopic spread in outcomes $x$ that cannot
be explained using statistical mixtures of {}``smaller'' states,
that is, states whose predictions give a smaller spread of outcome.
The issue of macroscopic quantum indeterminateness is fundamental
to quantum mechanics, in that any pure state can be written in terms
of eigenstates of any observable, and it is always the case that the
uncertainty principle will apply to prevent absolute predetermination
of another observable. Put another way, an eigenstate of momentum
when written in terms of position eigenstates will be a superposition
$|\psi\rangle=\sum_{i}c_{i}|x_i\rangle$ of a macroscopic --- in fact infinite --- 
range of position eigenstates $|x\rangle$.

In terms of Schrödinger's concern, we are left to question the real
existence of macroscopic quantum indeterminateness, since this would imply
a superposition of eigenstates with an inherently macroscopic range
of prediction of $x$. Following \cite{previous}, this is still a paradox. We consider two regions of outcome
(denoted $\pm1$) that are macroscopically separated, and denote the
region of intermediate outcomes by $0$, as shown in Figure 1. The
mixture $\rho=P_{1}\rho_{1}+P_{2}\rho_{2}$, where $\rho_{1}$ encompasses
outcomes $x<x_{2}$ and $\rho_{2}$ encompasses outcomes $x>x_{1}$
($P_{1/2}$ are probabilities), imposes a {}``macroscopic reality'', in the sense  that the system can be interpreted to be in possibly one (but
never both) of two macroscopically - separated regimes. The macroscopic
superpositions defy this assertion.

\begin{figure}
\begin{centering}\includegraphics[width=5.5cm,keepaspectratio]{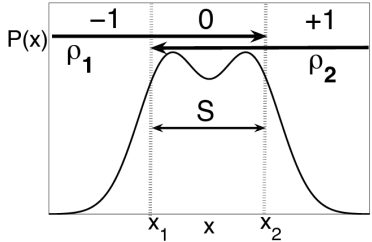}\par\end{centering}

\caption{Consider three regions of outcome $\pm1$, $0$ for measurement $\hat{x}$.
Density operator $\rho_{1}$ encompasses outcomes $x<x_{2}$ and $\rho_{2}$
encompasses outcomes $x>x_{1}$. }
\end{figure}

We present a unified approach for constructing criteria for macroscopic
superpositions and EPR entanglement. We first review some experimental
signatures\cite{previous} for determining the extent of {}``quantum
fuzziness''. These signatures are based on the use of quantum uncertainty
relations. Next, we show how one can easily construct from single-system
uncertainty relations new such signatures that apply to bipartite
entangled systems. These new signatures result by simply substituting
\emph{one} of the variances of the original uncertainty relation with
the variance of an inferred observable. Finally we show that the
simple further amendment of the uncertainty relations so that all
variances are replaced by inferred variances will result in criteria
for the EPR paradox\cite{epr}.

\section{Macroscopic realism, local realism and the completeness of quantum
mechanics}

The assumption we seek to test is \emph{macroscopic realism} (MR)\cite{leggett} --- 
that physical systems can always be described at any given time as being in one or other
of two macroscopically distinct states. This can (in principle) coexist
with a lack of such realism at the microscopic level.

EPR\cite{epr} argued against the completeness of quantum mechanics --- the notion that quantum mechanics is a complete theory in the sense
that there are no further facts about physical systems which are not
captured by a quantum description. In particular, quantum observables
obey uncertainty relations and the assumption of completeness implies
that the values of those observables are not defined beyond that precision.
EPR showed how this assumption of completeness of quantum mechanics
clashed with that of  \emph{local realism} (LR).

This assumption of the completeness of quantum mechanics does not
seem a priori to clash with MR --- an argument could be made that the
uncertainty principle imposes only a microscopic limitation on the
predetermination of observables. We show that this could be
a misleading argument, in that quantum mechanics \emph{predicts} the
existence of eigenstates of an observable (this observable is said
to be \emph{squeezed}) and thus implies infinite spreads in {}``quantum
fuzziness'', for conjugate observables. This prediction we wish to test.

\section{Criteria for S-scopic superpositions}

\textbf{Continuous variable case}: We consider a system $A$ for which
an observable $\hat{x}$ displays a macroscopic range of values. We
denote by $\hat{p}$ the observable conjugate to $\hat{x}$, so that
(in appropriate units) $\Delta^{2}x\Delta^{2}p\geq1$.

Leggett and Garg\cite{leggett} defined \emph{macroscopic realism}
(MR) as the assumption: "A macroscopic system with two or more macroscopically
distinct states available to it will at all times \emph{be} in one
or the other of these states". If we do not want to restrict a priori
what states are available to the system, we must assume that all possible
superpositions of eigenstates of $\hat{x}$ are available. If two
states each localized around macroscopically distinct values of $x$
indicate two macroscopically distinct states, then each (pure) quantum
state allowed by MR can only have a microscopic (or non-macroscopic)
range of outcomes.

In applying MR to situations where more than two states are available, we
thus postulate that MR asserts the system to be describable as a statistical
mixture of states $\rho^{(S)}_i$, each of which predicts a small (non-macroscopic)
spread of outcomes $x$ for $\hat{x}$. We now assume that the {}``states''
are \emph{quantum} states, and call this premise \emph{macroscopic
quantum} \emph{realism}. In this case, denoting the spread in the
prediction for $x$ for the state $\rho^{(S)}_i$ by $S$, we can write the density
matrix as
\begin{equation}
\rho=\sum_{i}P_{i}\rho_{i}^{(S)}\label{eq:rhos}
\end{equation}
 Here $\sum_{i}P_{i}=1$ and for each $\rho_{i}^{(S)}$, $|x_{1}-x_{2}|\leq S$
for all values of outcomes $x_{1}$ and $x_{2}$ which have zero probability. This assumption leads \cite{previous} to constraints on the minimum
fuzziness in the conjugate observable $p$. Specifically, it follows,
since each $\rho_{i}^{(S)}$ is itself a quantum state and since the
variance predicted by a mixture cannot be less than the average of
the variances of its components, that $\Delta^{2}p\geq\frac{4}{S^{2}}$.

\begin{figure}
\begin{centering}\includegraphics[width=5.5cm,keepaspectratio]{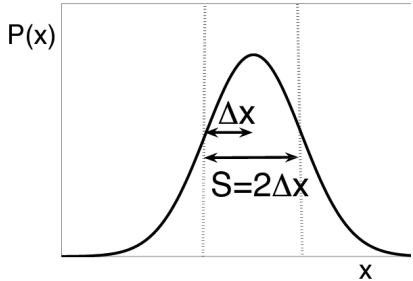}\par\end{centering}

\caption{Squeezed states predict a Gaussian distrbution for $x$ with variance
$\Delta x=e^{r}$. The measurement of a $\Delta p$ would imply superpositions
of $|x\rangle$ that have a range (or size) $S$ where $S>2/\Delta p$.
For the squeezed state, $S>2\Delta x$.}
\end{figure}

The experimental observation of squeezing in $p$ such that $\Delta p<2/S$
therefore implies the failure of mixtures of quantum states that can only have
a spread in their prediction for $x$ of $S$ or less. Thus necessarily
the system exists with some probability in a pure superposition state
of spread, or size, $S$ where
 \begin{equation}
S>2/\Delta p
\end{equation}
 The squeezed state\cite{sqst} $ $$|\psi\rangle=e^{r(a^2-a^{\dagger2})}\left|0\right\rangle $
($a$ is the boson operator for a field mode at $A$ and $\left|0\right\rangle $
is the vacuum state) is the simplest model for squeezed variances,
defined as $\Delta p<1$ (Fig. 2). Here measurements are: $\hat{x}=(a^{\dagger}+a)$,
$\hat{p}=i(a^{\dagger}-a)$. The squeezed state predicts
$\Delta^{2}x=\sigma=e^{2r}$, so that $x$ has eventually a macroscopic
quantum indeterminacy, while $p$ is squeezed, so that $\Delta^{2}p=1/\sigma=e^{-2r}$.
Experiments\cite{sqexp,spin sq,spin sq 2} using optical fields have
confirmed the existence of squeezed states. Values reported are of
order $\Delta p=0.4$, to confirm a quantum superposition of
eigenstates $|x\rangle$ with $S=4$,
which is twice that of the coherent state.

\textbf{Discrete case}: We present new criteria for the extent of
quantum indeterminateness for spin states with discrete outcomes. We use
$\Delta J_{X}\Delta J_{Y}\geq|\langle J_{Z}\rangle|/2$, where $J_x$, $J_Y$, $J_Z$ are angular momentum observables.  Suppose $\rho$
to be a mixture of superpositions of the eigenstates of $J_{X}$ that
have an extent $S$ or less. This leads to the constraint $\Delta J_{Y}\geq|\langle J_{Z}\rangle|/S$.
Thus if we measure a value $\Delta J_{Y}$ we can infer existence
of superpositions of size $S$ where \begin{equation}
S>|\langle J_{Z}\rangle|/\Delta J_{Y}\end{equation}
 The inequality is interesting in that the bound $|\langle J_{Z}\rangle|$
itself is not intrinsically restricted in size. This means that it
is possible to deduce existence of superpositions of spin eigenstates
which have a macroscopic extent in the indeterminateness, even if this
extent is small relative to the quantum limit itself.

One example is the observation of squeezing in {}``spin'' observables
constructed via the Schwinger formalism. We define 
$J_{X}^{A}=(a_{-}a_{+}^{\dagger}+a_{-}^{\dagger}a_{+})/2$, 
$J_{Y}^{A}=(a_{-}a_{+}^{\dagger}-a_{-}^{\dagger}a_{+})/2i$,
$J_{Z}^{A}=(a_{+}^{\dagger}a_{+}-a_{-}^{\dagger}a_{-})/2$,
where $a_{\pm}$ are boson operators for field
modes. The physical measurements are of photon number differences,
the $J_{X}$ and $J_{Y}$ measurements being performed by first combining
the fields with appropriate phase shifts. Thus, we define $a_{X\pm}=(a_{+}\pm a_{-})/\sqrt{2}$
and $a_{Y\pm}=(a_{+}\mp ia_{-})/\sqrt{2}$ to get $J_{X}=(a_{X+}^{\dagger}a_{X+}-a_{X-}^{\dagger}a_{X-})/2$
and $J_{Y}=(a_{Y+}^{\dagger}a_{Y+}-a_{Y-}^{\dagger}a_{Y-})/2$ . Squeezing
of spin variables for the macroscopic regime where outcomes become
effectively continuous has been observed, in experiments\cite{spinsq,spinsq2,spinpol,elizcold}
based on polarisation and atomic-spin squeezing.

\section{Criteria for S-scopic superpositions in bipartite systems}

\textbf{Continuous variable case}: We consider two subsystems $A$
and $B$, and define observables $x$, $p$ for $A$, and $x^{B}$,
$p^{B}$ for $B$, where $\Delta x^{B}\Delta p^{B}\geq1$ . We derive
an uncertainty relation that will be useful in deriving signatures
for superpositions of entangled systems.

\textbf{Theorem} 1: For any quantum state \begin{equation}
\Delta x\Delta_{inf}p\geq1\end{equation}
 We define the average variance in the inference of $p$ given a measurement
$\hat{O}^{B}$ at $B$ as $\Delta_{inf}^{2}p=\sum_{O^{B}}P(O^{B})\Delta^{2}(p|O^{B})$:
$\Delta^{2}(p|O^{B})$ is the variance of the conditional distribution
$P(p|O^{B})$ and $P(O^{B})$ is the probability of $O^{B}$, the
result for observable $\hat{O}^{B}$. In general, where we have a quantum
uncertainty relation of type $\Delta O_{1}\Delta O_{2}\geq|\langle[O_{1},O_{2}]\rangle|/2$,
or $\sum_{I}\Delta^{2}O_{I}\geq D$, we can construct another quantum
relation that applies to bipartite systems by substituting \emph{one}
of the variances, $\Delta^{2}O$ say, for the system $A$, with the
variance $\Delta_{inf}^{2}O$ of the inferred value for the observable $\hat{O}$.$ $

\textbf{Proof}: The variance $\Delta^{2}x$ is calculable from the
density operator for $A$ which is $\rho^{A}=Tr_{B}\rho=\sum_{O^{B}}P(O^{B})\rho_{O^{B}}^{B}$
where $\rho_{O^{B}}^{B}$ is the reduced state of $A$ conditional
on the result $O^{B}$ for the measurement $\hat{O}^B$ at $B$. We thus get
$\Delta^{2}x\geq\sum_{O^{B}}P(O^{B})\Delta_{O^{B}}^{2}(x|O^{B})$,
since the variance of a mixture can't be less than the average of
the variances of its components. Here we denote $\Delta_{O^{B}}^{2}(x|O^{B})$
as the variance of the conditional $P(x|O^{B})$. $ $Now using the
Cauchy Schwarz inequality
\begin{eqnarray}
\Delta^{2}x\Delta_{inf}^{2}p & \geq & \sum_{O^{B}}P(O^{B})\Delta^{2}(x|O^{B})\sum_{O^{B}}P(O^{B})\Delta^{2}(p|O^{B})\\*
 & \geq & \left[\sum_{O^{B}}P(O^{B})\Delta(x|O^{B})\Delta(p|O^{B})\right]^{2}\geq1\end{eqnarray}
 $ $Similar reasoning holds for the more general uncertainty relation
except that one uses $\Delta(O_{1}|O^{B})\Delta(O_{2}|O^{B})\geq|\langle C|O^{B}\rangle|/2$,
where $C=[O_{1},O_{2}]$ and
$\langle C|O^{B}\rangle$ denotes the average of $P(C|O^{B})$, and the fact
that in general  $\sum_{z}P(z)|\langle x|z\rangle|\geq\sum_{z}P(z)\langle x|z\rangle=\sum_{z}P(z)\sum_{x}xP(x|z)=\langle x\rangle$.
The result for the sums of variances can be proved in a similar fashion.

The assumption that $\rho$ can be expressed as a mixture of \emph{only}
S-scopic superpositions of  $|x\rangle$ will imply, following the logic outlined
in Section 3, the constraint $\Delta_{inf}p\geq2/S$. The observation
of a $\Delta_{inf}p$ allows us to deduce the existence of a superposition of eigenstates $|x\rangle$ with a spread $S$, where \begin{equation}
S>2/\Delta_{inf}p\end{equation}
 An arbitrary amount of squeezing $\Delta_{inf}p$ $ $is predicted
for the two-mode squeezed state~\cite{cavesch,eprr} $|\psi\rangle=\sum_{n=0}^{\infty}c_{n}|n\rangle_{A}|n\rangle_{B}$,
where $c_{n}=tanh^{n}r/coshr$. Here $\Delta x=\sigma=cosh2r$ while
$\Delta_{inf}p=1/\cosh2r$. The inference variance $\Delta_{inf}p$
has been measured and recorded in experiments\cite{eprexp} that are
designed to test for the EPR paradox. Values as low as $\Delta_{inf}p\approx0.7$
have been achieved.

\textbf{Discrete case}: We now consider where spin measurements $J_{\theta}$
and $J_{\phi}^{B}$ can be performed. Application of Theorem 1 leads
to the following inequality satisfied by all such quantum systems:
$ $$\Delta J_{X}\Delta_{inf}J_{Y}\geq|\langle J_{Z}\rangle|/2$.
The observation of a certain inference variance $\Delta_{inf}J_{Y}$
will lead to the conclusion of a superpositions of eigenstates of
$J_{X}$ with spread \begin{equation}
S>|\langle J_{Z}\rangle|/\Delta_{inf}J_{Y}\end{equation}
 Measurements of $\Delta_{inf}J_{Y}$ have been reported by Bowen
et al\cite{bowen}.

\section{Criteria for the EPR paradox }

We consider quantum uncertainty relations for system $A$ of a bipartite
system. For example we may have $\Delta O_{1}\Delta O_{2}\geq|\langle[O_{1},O_{2}]\rangle|/2$
where $[O_{1},O_{2}]$ evaluates as another observable which we denote
$C$. Alternatively, we may have\cite{hof} $\sum_{i}\Delta^{2}O_{i}\geq D$
where $D$ is a constant. Because we have a second system $B$, we
can define the inferred variances $\Delta_{inf}^{2}O_{i}$. The following
result allows an immediate writing down of criteria to confirm EPR's
paradox\cite{epr}.

\textbf{Theorem} 2: Where we have such a quantum uncertainty relation that holds for all quantum states, we can substitute the
variances $\Delta^{2}O$ by average inference variances $\Delta_{inf}^{2}O$,
and the mean $|\langle C\rangle|$ by $|\langle C\rangle|_{inf}$, the average inference  of the modulus
of the mean as defined by $|\langle C\rangle|_{inf}=\sum_{O^{B}}P(O^{B})|\langle C|O^{B}\rangle|$,
where $\langle C|O^{B}\rangle$ is the mean of the conditional distribution
$P(C|O^{B})$. The resulting inequality is an {}``EPR inequality''
that if violated is a demonstration of the EPR paradox.

\textbf{Proof:} We follow the treatment given by EPR\cite{epr},
and the modifications\cite{eprr,mermin},  to conclude existence of an {}``element
of reality'' $\mu_{Oi}$ that predetermines the result of measurement
for observable $O_{i}$. The probability distribution for the prediction
of this element of reality is precisely that of the conditional
$P(O_{i}|O^{B})$ where $O^{B}$ is the result of a measurement performed at
$B$, to infer the value of $O_{i}$. EPR's local realism (LR) implies
a joint probability distribution $P(\lambda)$ for the $\mu_{i}$,
or for further underlying parameters. For the product of the inference
variances we get, assuming LR\begin{eqnarray}
\Delta_{inf}^{2}O_1\Delta_{inf}^{2}O_2 & = & \sum_{O_{1}^B}P(O_{1}^{B})\Delta^{2}(O_{1}|O_{1}^{B})\sum_{O_{2}^B}P(O_{2}^{B})\Delta^{2}(O_{2}|O_{2}^{B})\\
 & \geq & [\sum_{\lambda}P(\lambda)\Delta(O_{1}|\lambda)\Delta(O_{2}|\lambda)]^{2}\\
 & \geq & \left|\sum_{\lambda}P(\lambda)|\langle C|\lambda\rangle|/2\right|^{2}\geq |\langle C\rangle|_{inf}^{2}/4\end{eqnarray} and for the sum one obtains 
 \begin{eqnarray}
 \Delta_{inf}^{2}(O_{1})+\Delta_{inf}^{2}(O_{2}) & = & \sum_{O_{1}^B}P(O_{1}^{B})\Delta^{2}(O_{1}|O_{1}^{B})+\sum_{O_{2}^B}P(O_{2}^{B})\Delta^{2}(O_{2}|O_{2}^{B})\nonumber \\
 & = & \sum_{\lambda}P(\lambda)[\Delta^{2}(O_{1}|\lambda)+\Delta^{2}(O_{2}|\lambda)]\geq D
 \end{eqnarray}
  We
have used\cite{eprr} that if the {}``elements of reality''
can be written as quantum states, then the variances predicted by
the elements of reality $\lambda$ must satisfy the quantum uncertainty
relations. This leads to the result (11), once it is realised that increasing
the number of variables $\lambda$ can only decrease the {\small average}
modulus of the mean.The violation of (11) or (12) thus implies inconsistency of LR with the completeness of quantum mechanics, that the underlying states symbolized by the elements of reality can be quantum states.

The {}``EPR inequalities'' 
 \begin{equation}
\Delta_{inf}^{2}x\Delta_{inf}^{2}p\geq1,\,\,\,\,\Delta_{inf}J_{X}\Delta_{inf}J_{Y}\geq|\langle J_{Z}\rangle_{inf}|/2\end{equation}
 (the latter implies the further EPR inequality\cite{bowen} $\Delta_{inf}J_{X}\Delta_{inf}J_{Y}\geq|\langle J_{Z}\rangle|/2$)
have been derived previously\cite{eprr} and in some cases used
to demonstrate an EPR paradox\cite{eprexp,bowen}. One can also use
Theorem 2 to derive EPR inequalities from uncertainty relations involving
sums of variances, so that for example $\Delta^{2}J_{X}+\Delta^{2}J_{Y}+\Delta^{2}J_{Z}\geq j/2$
as used by Hoffmann et al\cite{hof} leads to the EPR inequality
$\Delta_{inf}^{2}J_{X}+\Delta_{inf}^{2}J_{Y}+\Delta_{inf}^{2}J_{Z}\geq j$/2.

\section{Conclusion}

The criteria we have derived are based on the assumption that the systems can be described as  mixtures of underlying \emph{quantum} states, which therefore satisfy uncertainty relations. This means that the criteria can be
viewed in a unified way as conditions for demonstration of general
EPR-type paradoxes. In the case of the criteria for macroscopic superpositions,
we \emph{assume macroscopic realism} (MR) to infer that the system
be described as probabilistic mixture of states with a \emph{microscopic
lack of predetermination} only. The assumption that these underlying
states be \emph{quantum} states leads to our inequalities. An experimental
violation of the inequalities confirms existence of macroscopic superpositions,
but does not falsify macroscopic realism itself, since one may propose
alternative theories in which the underlying states are \emph{not
quantum} states. Hence we have extended the EPR paradox to demonstrate
an inconsistency between completeness of quantum mechanics and \emph{macroscopic
realism}.


\begin{thebibliography}{10}
\bibitem{sch}E. Schrödinger, Naturwissenschaften \textbf{23}, 807
(1935). 

\bibitem{leggett}A. J. Leggett and A. Garg, Phys. Rev. Lett. 54,857
(1985).

\bibitem{previous}E. G. Cavalcanti and M. D. Reid, Phys. Rev. Lett.,
\textbf{97}, 170405 (2006).

\bibitem{epr}A. Einstein, B. Podolsky and N. Rosen, Phys. Rev. \textbf{47},
777 (1935).

\bibitem{sqst}H. P. Yuen, Phys. Rev. A\textbf{13}, 2226 (1976).

\bibitem{sqexp}S. Suzuki S , H. Yonezawa, F. Kannari, M. Sasaki,
A. Furusawa, App. Phys. Lett. 89, 061116 (2006).

\bibitem{spinsq}J. F. Corney, P. D. Drummond, J. Heersink, V. Josse,
G. Leuchs, and U. L. Andersen, Phys. Rev. Lett. \textbf{97}, 023606
(2006) .

\bibitem{spinsq2}W. P. Bowen, R. Schnabel, H. A. Bachor and P. K.
Lam, Phys. Rev. Lett. \textbf{88}, 203601 (2002).

\bibitem{elizcold}V. Josse, A. Dantan, L. Vernac, A. Bramati, M.
Pinard, and E. Giacobino, Phys. Rev. Lett. \textbf{91}, 103601 (2003).

\bibitem{spinpol}B. Julsgaard, A. Kozhekin and E. S. Polzik, Nature
\textbf{413}, 400 (2001).

\bibitem{cavesch}C. M. Caves and B. L. Schumaker, Phys. Rev. A \textbf{31},
3068 (1985).

\bibitem{eprr}M. D. Reid, Phys. Rev. A \textbf{40}, 913 (1989); quant-ph/0103142.

\bibitem{eprexp}Z. Y. Ou et al, Phys. Rev. Lett. \textbf{68},
3663 (1992). Yun Zhang et al, Phys. Rev. A \textbf{62}, 023813 (2000).
C. Silberhorn et al, Phys. Rev. Lett. \textbf{86}, 4267 (2001). W.
P. Bowen et al, Phys. Rev. Lett. \textbf{90}, 043601 (2003). J. C.
Howell et al, Phys. Rev. Lett. \textbf{92}, 210403 (2004).

\bibitem{bowen}W. P. Bowen, N.~Treps, R.~Schnabel, P . K. Lam,
Phys. Rev. Lett. \textbf{89}, 253601 (2002).

\bibitem{hof}H. F. Hofmann and S. Takeuchi, Phys. Rev. A \textbf{68},
032103 (2003). 

\bibitem{mermin}Mermin, N. D., Physics Today \textbf{43}, 9 (1990).
\end{thebibliography}
\end{document}